\documentclass[prd,nofootinbib,preprint,superscriptaddress,twocolumn,10pt]{revtex4}
\pdfoutput=1
\usepackage{amsmath,amssymb}
\usepackage{epsfig}
\usepackage{graphicx}
\usepackage[usenames,dvipsnames]{color}
\usepackage{subfigure}
\usepackage{slashed}
\usepackage{comment}
\usepackage[colorlinks,citecolor=blue]{hyperref}
\usepackage{color}

\usepackage{multirow}
\usepackage{booktabs}

\begin{document}

\title{Light thermal dark matter models in the light of DAMIC‑M 2025 constraints}

	%%%%%%%%%   Authors   %%%%%%%%%%%%		
\author{{Debasish Borah}}
\email{dborah@iitg.ac.in}
\affiliation{Department of Physics, Indian Institute of Technology Guwahati, Assam 781039, India}

    \author{Satyabrata Mahapatra}
	\email{satyabrata@iitgoa.ac.in}
    \affiliation{School of Physical Sciences, Indian Institute of Technology Goa, Ponda-403401, Goa, India }
	%\affiliation{Department of Physics and Institute of Basic Science, Sungkyunkwan University, Suwon 16419, Korea}
		
	\author{Narendra Sahu}
	\email{nsahu@phy.iith.ac.in}
	\affiliation{Department of Physics, Indian Institute of Technology Hyderabad, Kandi, Sangareddy 502285, Telangana, India}
	
	\author{Vicky Singh Thounaojam}
    \email{ph22resch01004@iith.ac.in}
	\affiliation{Department of Physics, Indian Institute of Technology Hyderabad, Kandi, Sangareddy 502285, Telangana, India}
    
	\begin{abstract}
We study the viability of light thermal dark matter (DM) in sub-GeV mass range in view of the stringent new DAMIC‑M limits on DM–electron scattering. Considering a Dirac fermion singlet DM charged under a new Abelian gauge symmetry $U(1)$, we outline two possibilities: (i) family non-universal $U(1)$ gauge coupling with resonantly enhanced DM annihilation into standard model (SM) fermions and (ii) family universal dark $U(1)$ gauge symmetry where relic is set by DM annihilation into light gauge bosons. As an illustrative example of the first class of models, we consider a gauged $L_\mu-L_\tau$ extension of the SM having interesting detection prospects at several experiments. While both of these class of models lead to observed DM relic and consistency with DAMIC-M together with other experimental limits, the second class of models also lead to strong DM self-interactions, potentially solving the small-scale structure issues of cold dark matter. While a vast part of the parameter space in both the models is already ruled out, the current allowed region of parameter space can be further probed at ongoing or future experiments keeping the models testable.

%With a vast part of the parameter space in both the models being ruled out already, the currently allowed region of parameter space can be further probed at ongoing or future experiments keeping the models testable in near future.

%We present a comprehensive study of light dark matter (DM) in two distinct gauge symmetric frameworks in light of the stringent new DAMIC‑M limits on sub‑GeV dark matter–electron scattering. In the first scenario, we consider a gauged $L_\mu-L_\tau$ extension of the Standard Model (SM), where DM interacts with the SM predominantly via a $Z_{\mu\tau}$ boson with the latter being responsible for resonantly enhanced DM annihilation required for the desired thermal relic and a direct-detection rate in the correct ballpark due to one-loop suppressed kinetic mixing ($\varepsilon \sim g_{\mu\tau}/70$). In the second scenario, a generic dark $U(1)_X$ gauge extension of the SM is considered, which offers a greater flexibility via a free kinetic mixing parameter. However, requiring sufficient DM self-interaction to resolve the small-scale structure formation issues in astrophysics and relic density dominantly set by annihilation into dark gauge bosons significantly restrict the viable parameter space. We show that in both setups, one can satisfy DAMIC‑M’s electron-scattering constraints while reproducing the observed DM abundance consistent with other experimental and phenomenological constraints, including the recently updated constraints on muon $g-2$ applicable for $L_\mu-L_\tau$ scenario.

	\end{abstract}	
	\maketitle
	%\flushbottom
	
\noindent
\section{Introduction}\label{sec:Introduction}
The persistent null results from direct-detection experiments~\cite{XENON:2025vwd, LZ:2024zvo, PandaX:2024qfu} have shifted considerable attention in the dark matter (DM) community toward exploring the light (sub-GeV) dark matter regime~\cite{Balan:2024cmq, Cheek:2025nul, Krnjaic:2025noj, Dutta:2019fxn,Essig:2017kqs,Bondarenko:2019vrb,Adhikary:2024btd,Borah:2024yow}. While Weakly Interacting Massive Particles (WIMPs)~\cite{ Bertone:2004pz, Arcadi:2017kky,Arcadi:2024ukq} in the GeV--TeV scale had long stood as leading DM candidates, recent experimental advances employing ultra-low threshold detectors have notably improved sensitivity to lighter DM particles~\cite{XENON:2019gfn, CRESST:2019jnq, XENON:2024znc, PandaX-II:2021nsg, PandaX:2022xqx, SENSEI:2020dpa, SuperCDMS:2024yiv}. Among these, the DAMIC (Dark Matter in CCDs) collaboration, particularly through their most recent DAMIC-M results, has provided the most stringent constraints to date on DM-electron scattering for sub-GeV DM masses~\cite{DAMIC-M:2025luv}. This result, not only significantly improves the direct-detection sensitivity to previously unconstrained ranges but also exclude vast regions of parameter space for a wide class of hidden sector DM models, especially those where DM interactions with the SM particles arise through feeble mixing of a dark sector gauge boson with the photon. In these so-called "Dark Photon" or "Dark gauge-boson" scenarios, the cross-section for DM-electron scattering is highly sensitive to the strength of the kinetic mixing parameter making the latest DAMIC-M bounds crucially relevant for such models. 

Recently, sub-GeV DM models which are consistent with the latest DAMIC-M results are studied in Refs. \cite{Cheek:2025nul, Krnjaic:2025noj}. In~\cite{Cheek:2025nul}, the authors derive competitive limits from PandaX-4T S2-only data and surveys their effect on several generic DM models. On the other hand, in~\cite{Krnjaic:2025noj}, the authors perform a status update of thermal DM coupled to a generic dark photon, arguing that direct-detection results now robustly exclude the complex scalar DM scenario. In this work, we perform a detailed, multi-messenger analysis of two specific, well-motivated sub-GeV DM models; $U(1)_{L_\mu-L_\tau}$ and $U(1)_X$ to demonstrate the survival parameter space which not only satisfy the direct detection results from DAMIC-M, but also satisfy constraints from cosmology, astrophysics, and precision measurements like $(g-2)_\mu$. We show that, in these models, despite the stringent constraints from DAMIC-M, it is possible to achieve the observed dark matter relic density  with phenomenologically viable parameter space consistent with other experimental bounds.

%In this work, we revisit this paradigm in the context of two well-motivated models.
%\footnote{\DB{Footnote shifted to main text as it was too long.}}. Refs. \cite{Cheek:2025nul, Krnjaic:2025noj} studied generic light DM models which are consistent with DAMIC-M results. %Here we constrain two well-motivated light DM scenarios and compare DAMIC-M bounds with other relevant bounds like muon $g-2$, DM self-interactions. Ref.~\cite{Cheek:2025nul} derives competitive limits from PandaX-4T S2-only data and surveys their effect on several generic DM models. Ref.~\cite{Krnjaic:2025noj} performs a status update of thermal DM coupled to a generic dark photon, arguing that direct-detection results now robustly exclude the complex scalar DM scenario. Our work, in contrast, performs a detailed, multi-messenger analysis of two specific, well-motivated models; $U(1)_{L_\mu-L_\tau}$ model and a $U(1)_X$ model demonstrating concrete parameter spaces where they remain viable against a comprehensive set of constraints beyond direct detection such as from cosmology, astrophysics, and precision measurements like the muon $(g-2)$. We demonstrate that, despite the stringent constraints from DAMIC-M,  it remains possible to achieve the observed dark matter relic density  with phenomenologically viable parameter space consistent with other experimental bounds. 

The first scenario we consider is based on a gauged $L_\mu-L_\tau$ extension of the SM~\cite{Foot:1990mn,He:1990pn, He:1991qd} with family non-universal couplings, motivated not only by its theoretical appeal as an anomaly-free gauge symmetry but also by its experimental verifiability~\cite{Bauer:2018onh,Bernal:2025szh}. The $Z_{\mu\tau}$ gauge boson associated with this symmetry is being actively searched for at present and future colliders and fixed-target experiments. Furthermore, this framework is intimately connected to the longstanding discrepancy between the SM and measured value of the anomalous magnetic moment of the muon~\cite{Baek:2001kca, Ma:2001md}. While there exists no significant discrepancy between experimental data and updated SM predictions for muon $g-2$ \cite{Aliberti:2025beg}, the latest results from Fermilab~\cite{Muong-2:2025xyk} can be used to put an upper bound on the $L_\mu-L_\tau$ gauge coupling depending on the gauge boson mass. On the other hand, the flavor-dependent interactions of this model leads to a pronounced suppression of dark matter-electron scattering, as the dominant interactions are with muons and taus. The resulting kinetic mixing with the photon, $\epsilon \sim g_{\mu\tau}/70$, is sufficiently small such that, despite the strong DAMIC-M bounds, the correct relic abundance can be realized through resonant annihilation of DM via the $Z_{\mu\tau}$ portal.

The second scenario we analyze is a generic dark $U(1)_X$ gauge extension with family universal couplings via kinetic mixing of $U(1)_X$ with $U(1)_Y$ of the SM. This construction allows, in principle, more flexibility as the gauge kinetic mixing parameter is now unconstrained and can be treated as a free parameter. Due to the possibility of light gauge boson and relatively larger gauge coupling\footnote{Self-interacting DM was also studied in the context of gauged $L_\mu-L_\tau$ model by the authors of \cite{Kamada:2018zxi}. But for sub-GeV DM motivated from the recent DAMIC-M results, the experimental constraints on the model parameter space does not allow sufficient self-interactions.}, such a dark $U(1)_X$ extension also allows large DM self-interactions which can address small-scale structure formation issues in cold dark matter (CDM) scenarios such as core-cusp and too-big-to-fail problems~\cite{Spergel:1999mh,Tulin:2017ara,Bullock:2017xww}  . When the dual requirements of velocity-dependent DM self-interactions~\cite{Buckley:2009in,Feng:2009hw,Feng:2009mn,Loeb:2010gj,Bringmann:2016din,Kaplinghat:2013yxa,Tulin:2013teo}, and the observed relic density are imposed, viable parameter space gets tightly constrained. In particular, in the regime where dark matter annihilates predominantly into dark gauge bosons to achieve correct relic density, we show that achieving consistency with DAMIC-M and cosmological constraints is possible, but only within tightly restricted corners of the parameter space. 

%\DB{Can we comment on self-interacting DM in $L_\mu-L_\tau$ model \cite{Kamada:2018zxi} and the reason why we are not considering it here instead of using a generic $U(1)_X$? }{\textcolor{blue}{\textit{In Kamada's paper, for DM self interaction, the scalar that breaks the symmetry acts as the mediator and the gauge boson explains (g-2). DM relic is achieved via its annihilation to both gauge boson and scalar; thus more freedom but the viable DM mass lies in a range of few GeV to 100 GeV. In the current work, we do not discuss how $L_\mu-L_\tau$ gauge boson is getting its mass. And the viable range of coupling and gauge boson mediator mass we have, can not explain the DM self-interaction.}}}

By performing a detailed numerical analysis, we provide a comparative analysis of these two scenarios and identify viable ranges for the DM and mediator masses, couplings, and kinetic mixing where they not only evade the latest direct-detection limits but also satisfy correct relic density as well as all other existing experimental, cosmological and phenomenological constraints. 
The remainder of this paper is organized as follows. In Sec.~\ref{sec:LmuLtau}, we present the details of the $U(1)_{L_\mu-L_\tau}$ model, discussing its direct-detection prospects, contribution to the muon anomalous magnetic moment, relic density, and associated constraints, with particular emphasis on how the recent DAMIC-M limits can be consistently satisfied. In Sec.~\ref{sec:U1X}, we analyze a generic $U(1)_X$ extension, focusing on dark matter direct-detection, self-interactions, relic density, and the corresponding phenomenological limits, again highlighting the regions where DAMIC-M bounds can be accommodated along with other experimental and cosmological constraints. %A comparative discussion of viable parameter spaces is given together with the relevant experimental bounds. 
We finally summarize our results and conclude in Sec.~\ref{sec:conclusion}. Additional technical details, including expressions for self-interaction cross-sections, decay widths, and constraints from $N_{\rm eff}$, are provided in the Appendices.

\section{$\boldsymbol{U(1)_{L_{\mu}-L_{\tau}}}$ Model}\label{sec:LmuLtau}
The $U(1)_{L_\mu - L_\tau}$ gauge extension of the SM provides a simple yet well-motivated framework in which the difference of muon and tau lepton numbers is promoted to a local symmetry \cite{Foot:1990mn,He:1990pn, He:1991qd}. This construction is particularly attractive because it is anomaly-free without the need for additional fermions, and establishes a natural connection to neutrino flavor physics. While this model has received lots of attention as an appealing explanation of the previously significant anomaly of muon magnetic moment, in the present scenario, the motivation for this specific framework is primarily due to the suppressed DM-electron couplings that help evade stringent direct-detection bounds from DAMIC-M for sub-GeV DM.

In the minimal setup, the muon and tau leptons (together with their associated neutrinos) carry charges $+1$ and $-1$, respectively, under $U(1)_{L_\mu-L_\tau}$. We introduce a SM-singlet Dirac fermion $\chi$ as the dark matter candidate with $U(1)_{L_\mu-L_\tau}$ charge $q_\chi$, ensuring anomaly cancellation. The relevant interactions of $U(1)_{L_\mu-L_\tau}$ gauge boson with SM and SM are described by   
\begin{equation}
	\mathcal{L} \supset -g_{\mu\tau} (Z_{\mu \tau})_{\lambda}\Big(\sum_i q_{f_i}\, \overline{f_i}\gamma^\lambda f_i \Big) - q_\chi g_{\mu\tau}\, (Z_{\mu \tau})_{\lambda}\, \overline{\chi}\gamma^\lambda \chi , \nonumber 
\end{equation}
where $f_i=\{\mu,\tau,\nu_\mu,\nu_\tau\}$ and the gauge charges are assigned as $q_{f_i}=+1$ for $(\mu,\nu_\mu)$ and $q_{f_i}=-1$ for $(\tau,\nu_\tau)$. The gauge boson $Z_{\mu\tau}$ acquires its mass $M_{Z_{\mu\tau}}$ through the spontaneous breaking of $U(1)_{L_\mu-L_\tau}$ by an appropriate scalar field. Without any loss of generality we set $q_\chi=1/2$ which stabilizes the DM without requiring any additional discrete symmetries.

Here, it is important to emphasize that a kinetic mixing term between the SM hypercharge gauge group $U(1)_Y$ and the new $U(1)_{L_\mu-L_\tau}$ can, in general, appear in the Lagrangian, taking the form $\frac{\epsilon}{2} B_{\alpha\beta} Y^{\alpha\beta}$. Here, $B_{\alpha\beta}=\partial_\alpha X_\beta - \partial_\beta X_\alpha$ and $Y_{\alpha\beta}$ denote the field strength tensors of $U(1)_{L_\mu-L_\tau}$ and $U(1)_Y$, respectively, while $\epsilon$ is the mixing parameter. Even if this term is absent at the tree level, it is inevitably generated radiatively at one loop through particles carrying charges under both gauge groups. The one-loop mixing is given by\cite{Kamada:2015era},  
\begin{equation}
	\Pi(q^{2})=\frac{8 e g_{\mu\tau}}{4 \pi^2} \int_0^1 dx~x(1-x)\ln \left(\frac{m_\tau^2-x(1-x)q^2}{m_\mu^2-x(1-x)q^2}\right) , \nonumber 
\end{equation}
which in the limit $q \ll m_\mu$ reduces to
\begin{equation}
	\epsilon_A \simeq - \frac{e g_{\mu\tau}}{12 \pi^2} \ln\!\left(\frac{m_\tau^2}{m_\mu^2} \right) \simeq -\frac{g_{\mu\tau}}{70}.
\end{equation}

\subsubsection*{Muon Anomalous Magnetic Moment}
The longstanding discrepancy between the experimental measurement and the SM prediction of the muon anomalous magnetic moment $a_\mu = (g-2)_\mu/2$ has motivated numerous new physics scenarios. Although the latest lattice-QCD based SM prediction \cite{Aliberti:2025beg} indicates that the tension may no longer be statistically significant, the average of the recent Fermilab measurement \cite{Muong-2:2025xyk} with updated SM inputs yields  
\[
\Delta a_\mu = (39 \pm 64)\times 10^{-11},
\]
corresponding to a $1\sigma$ deviation. In the $U(1)_{L_\mu-L_\tau}$ framework, the new gauge boson $Z_{\mu\tau}$ contributes to this observable at one loop \cite{Baek:2001kca,Lynch:2001zr,Lindner:2016bgg},
\begin{equation}
	\Delta a_\mu=\frac{g_{\mu\tau}^4}{8\pi^2} \int_0^1 dx \,\frac{2 m_\mu^2 x^2(1-x)}{x^2 m_\mu^2+(1-x) M_{Z_{\mu\tau}}^2}.
\end{equation}
In the present analysis, rather than interpreting this deviation as a genuine anomaly, we employ the Fermilab measurement as an experimental constraint on the parameter space of the model, specifically on the gauge coupling $g_{\mu\tau}$ and the mass of the mediator $M_{Z_{\mu\tau}}$.

% \subsubsection*{Muon Anomalous Magnetic Moment}
% The muon anomalous magnetic moment, $a_\mu = (g-2)_\mu/2$, quantifies the deviation of the muon’s magnetic dipole moment from the Standard Model prediction.
% Although lattice-QCD results suggest that the discrepancy with experiment may be statistically insignificant \cite{Aliberti:2025beg}, the latest Fermilab $g-2$ measurement \cite{Muong-2:2025xyk}, combined with the updated SM prediction \cite{Aliberti:2025beg}, yields a new world average showing a $1\sigma$ difference, $\Delta a_\mu = (39 \pm 64) \times 10^{-11}$.
% In the present model, this deviation arises from the contribution of the gauge boson $Z_{\mu\tau}$, with the muon $g-2$ correction given by \cite{Baek:2001kca,Lynch:2001zr,Lindner:2016bgg}
% \begin{equation}
%     \Delta a_\mu=\frac{g_{\mu\tau}^4}{8 \pi^2} \int_0^1 dx~\frac{2 m_\mu^2 x^2(1-x)}{x^2 m_\mu^2+(1-x)M_X^2}
% \end{equation}

\subsubsection*{Dark Matter Relic Density}

For sub-GeV dark matter, achieving the observed relic density via the conventional freeze-out mechanism is typically challenging, since the annihilation cross-section is suppressed and thus insufficient to deplete the dark matter population. In fact, for DM interactions typically in the WIMP ballpark, the requirement of DM not overclosing the Universe leads to a lower bound on its mass, around a few GeV \cite{Lee:1977ua, Kolb:1985nn}. However, with inclusion of light mediators or additional particles it is possible to realise light thermal DM, as pointed out in several works \cite{Pospelov:2007mp, DAgnolo:2015ujb, Berlin:2017ftj, DAgnolo:2020mpt, Herms:2022nhd, Jaramillo:2022mos, Borah:2024yow}. In the present $U(1)_{L_\mu-L_\tau}$ setup, the process $\chi \overline{\chi} \to {f_i} \overline{{f_i}}$ mediated by the gauge boson $Z_{\mu\tau}$ governs freeze-out, and the tree-level annihilation cross-section takes the form
\begin{equation}
	\sigma(s) = \frac{N_{f_i}~g_{\mu\tau}^4}{24 \pi} \sqrt{\frac{s-4m_{f_i}^2}{s-4m_\chi^2}}
	\frac{s^2+2(m_\chi^2+m_{f_i}^2)s+4m_{f_i}^2 m_\chi^2}{s\big[(s-M_{Z_{\mu\tau}}^2)^2+\Gamma_{Z_{\mu\tau}}^2 M_{Z_{\mu\tau}}^2\big]} ,
\end{equation}
where $\Gamma_{Z_{\mu\tau}}$ is the total decay width of $Z_{\mu\tau}$ (see Appendix~\ref{app::decay_width}) and $N_{f_i}=2$ for charged leptons, while $N_{f_i}=1$ for neutral leptons. This factor arises due to equal contributions from both left-handed and right-handed chiral components of charged leptons, while for the neutral leptons, there is contribution only from left-handed component. The thermal averaged annihilation cross-section is given by:
\begin{equation}
	\langle \sigma v \rangle \simeq \frac{N_{f_i}~g_{\mu\tau}^4}{2\pi}\,\frac{(2 m_\chi^2+m_{f_i}^2)}{\left(4m_\chi^2-M_{Z_{\mu\tau}}^2\right)^2}\sqrt{1-\frac{m_{f_i}^2}{m_\chi^2}} + \mathcal{O}(v_\chi^2),
	\label{eq:sigmaVLmLt}
\end{equation}

%This is a specific feature of the model with a universal coupling for each chiral components to the gauge boson.

For $m_\chi,\,M_{Z_{\mu\tau}} < m_\mu$, the only kinematically open annihilation channels are into neutrinos, making $\chi \chi \rightarrow \bar{\nu_i}\nu_i, i \equiv \mu, \tau$ the dominant contribution. Away from resonance, the cross-section is too small to account for the observed relic density. However, the correct abundance can be achieved near the resonance condition, $M_{Z_{\mu\tau}} \simeq 2 m_\chi$, where annihilation is resonantly enhanced. This allows the dark matter population to freeze out with the appropriate relic density without requiring unnaturally large gauge couplings, which are otherwise severely constrained by various experimental searches. While DM can also annihilate into $e^- e^+$ final states via one-loop kinetic mixing, it remains suppressed compared to the neutrino final states mentioned above. The same kinetic mixing is also responsible for keeping DM-electron scattering rates below the stringent upper limits set by experiments like DAMIC-M.

\subsubsection*{Dark Matter direct detection}
Direct-detection experiments search for dark matter through its elastic scattering with detector targets, where the interaction deposits a tiny amount of measurable energy in the form of ionization, scintillation or phonon excitations. For dark matter heavier than a few GeV, the dominant channel is scattering off nuclei, producing nuclear recoils. However, for light dark matter in the MeV–GeV range, nuclear recoils are typically below threshold, and the most sensitive probes come instead from DM–electron scattering.  

In the present $U(1)_{L_\mu-L_\tau}$ framework, even though DM-electron scattering is suppressed due one-loop kinetic mixing, it can still be constrained from low-threshold experiments such as DAMIC-M and DarkSide-50. The cross-section for elastic DM-electron scattering is given by \cite{Essig:2015cda} 
\begin{equation}
    \sigma_{\chi e} = \frac{\mu_{\chi e}^2}{\pi} 
    \frac{\epsilon_A^2 e^2 g_{\mu\tau}^2}{\left(M_{Z_{\mu\tau}}^2 + \alpha^2 m_e^2\right)^2} ,
    \label{eq:DDLmLt}
\end{equation}
where $\mu_{\chi e}$ is the reduced mass of the DM–electron system.  This suppression of DM-electron coupling by $\epsilon_A$ allows a large portions of the parameter space consistent with relic density requirements to remain viable even under the most recent DAMIC-M and DarkSide-50 constraints.

% Dark matter direct detection aims to measure rare interactions with ordinary matter via the small energy deposited in a target, observable as light, ionization, or vibrations. Nuclear recoils are suited for heavier dark matter, whereas electron scattering is crucial for light (sub-GeV) dark matter that cannot produce detectable nuclear recoils. For dark matter masses between $1\ \mathrm{MeV}$ and $1\ \mathrm{GeV}$, the strongest limits arise from DM–electron scattering, with the cross section given by \cite{Essig:2015cda}

% \begin{equation}
%     \sigma (\chi e \leftrightarrow \chi e)=\frac{\mu_{\chi e}^2}{\pi}~\frac{\epsilon_A^2 e^2 g_{\mu\tau}^2}{(M_{Z_{\mu\tau}}^2+\alpha^2 m_e^2)^2}
%     \label{eq:DDLmLt}
% \end{equation}
% where $\mu_{\chi e}$ is the reduced mass of the dark matter electron system.

\begin{figure}[h]
    \centering
    \includegraphics[width=0.43\textwidth]{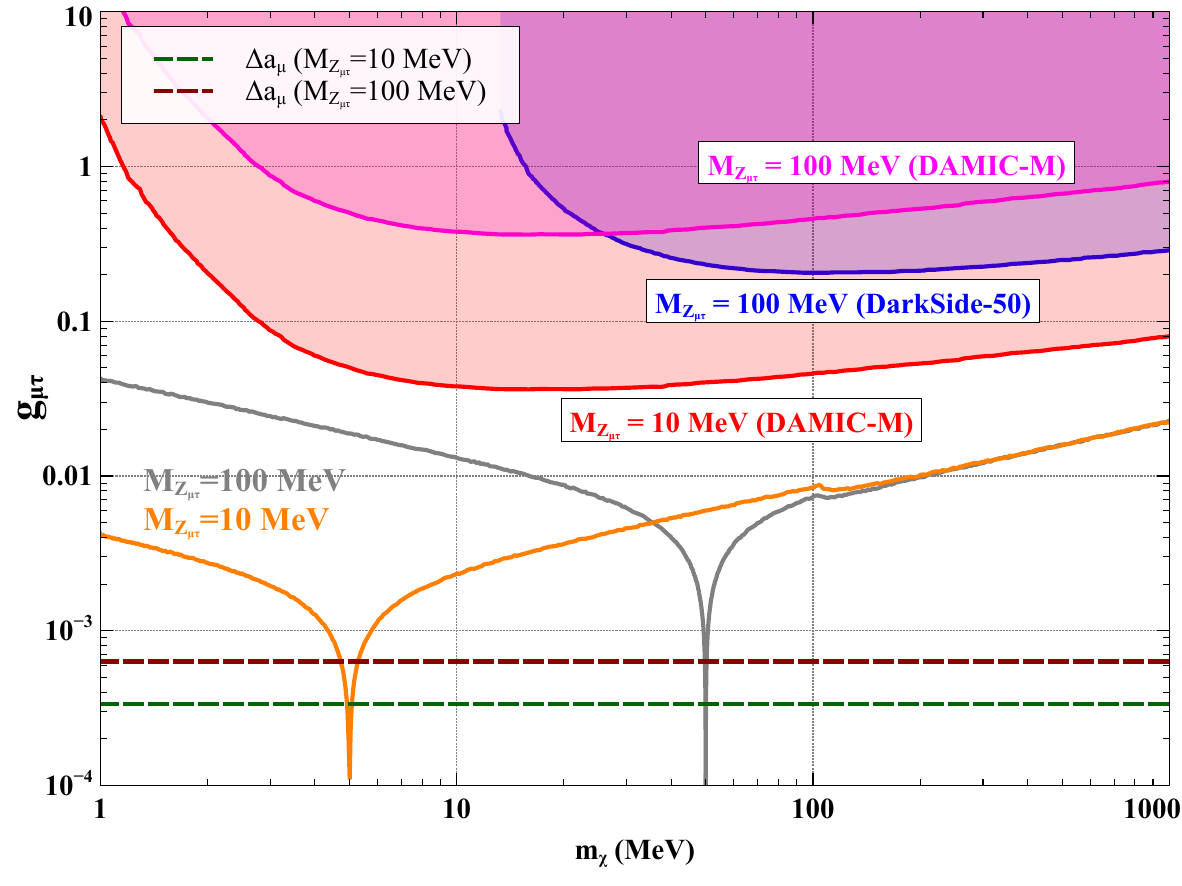}
    \caption{Relic density favored parameter space in the $(g_{\mu\tau}, m_\chi)$ plane for $M_{Z_{\mu\tau}} = 10$ MeV (orange) and $100$ MeV (gray). Direct-detection constraints from DAMIC-M and DarkSide-50, as well as upper limits from $(g-2)_\mu$, are overlaid.
    %The correct relic density points for two values of $M_X$ = 10 MeV (orange) and 100 MeV(grey) in the plane of $g_{\mu\tau}$ and $m_\chi$ with the direct detection limits from DAMIC-M and DarkSide-50. Moreover, the upper bound on $g_{\mu\tau}$ is also shown for the two values of $M_X$ as dashed lines: green (10 MeV) and purple (100 MeV).
    }
    \label{fig:ddrelic}
\end{figure}

In Fig.~\ref{fig:ddrelic}, we showcase the relic density favored parameter space in the $(g_{\mu\tau}, m_\chi)$ plane for two benchmark mediator masses, $M_{Z_{\mu\tau}} = 10$ MeV (orange) and $100$ MeV (gray). Direct-detection limits from DAMIC-M \cite{DAMIC-M:2025luv} and DarkSide-50 \cite{DarkSide:2022knj} are overlaid, together with the exclusion curve from the muon anomalous magnetic moment $(g-2)_\mu$. A sharp dip in the correct relic density satisfying contour is observed around $m_\chi \simeq M_{Z_{\mu\tau}}/2$, corresponding to the resonance which enhances the cross-section, lowering the required coupling. In this setup, the present $(g-2)_\mu$ constraints are more restrictive than those from direct-detection, such that only parameter points lying in or near the resonant regime remain viable after these constraints are applied. 

\subsubsection*{Indirect detection and CMB constraints} 
Thermal DM with masses below $\sim 20~\mathrm{GeV}$ is subject to severe constraints from the Cosmic Microwave Background (CMB) and indirect-detection searches, as the annihilation cross-sections excluded by these observations lie well below the canonical thermal value required for freeze-out. In this work, we examine the most stringent of these limits by focusing on DM annihilation into $e^+e^-$ final states, which can be used to scrutinize the viable parameter space of the $U(1)_{L_\mu-L_\tau}$ model. Even though the annihilation into electrons is suppressed in this framework due to the loop-induced kinetic mixing parameter $\epsilon_A \sim g_{\mu\tau}/70$, but the resulting bounds still remain very significant.  
\begin{figure}[h]
    \centering
    \includegraphics[width=0.43\textwidth]{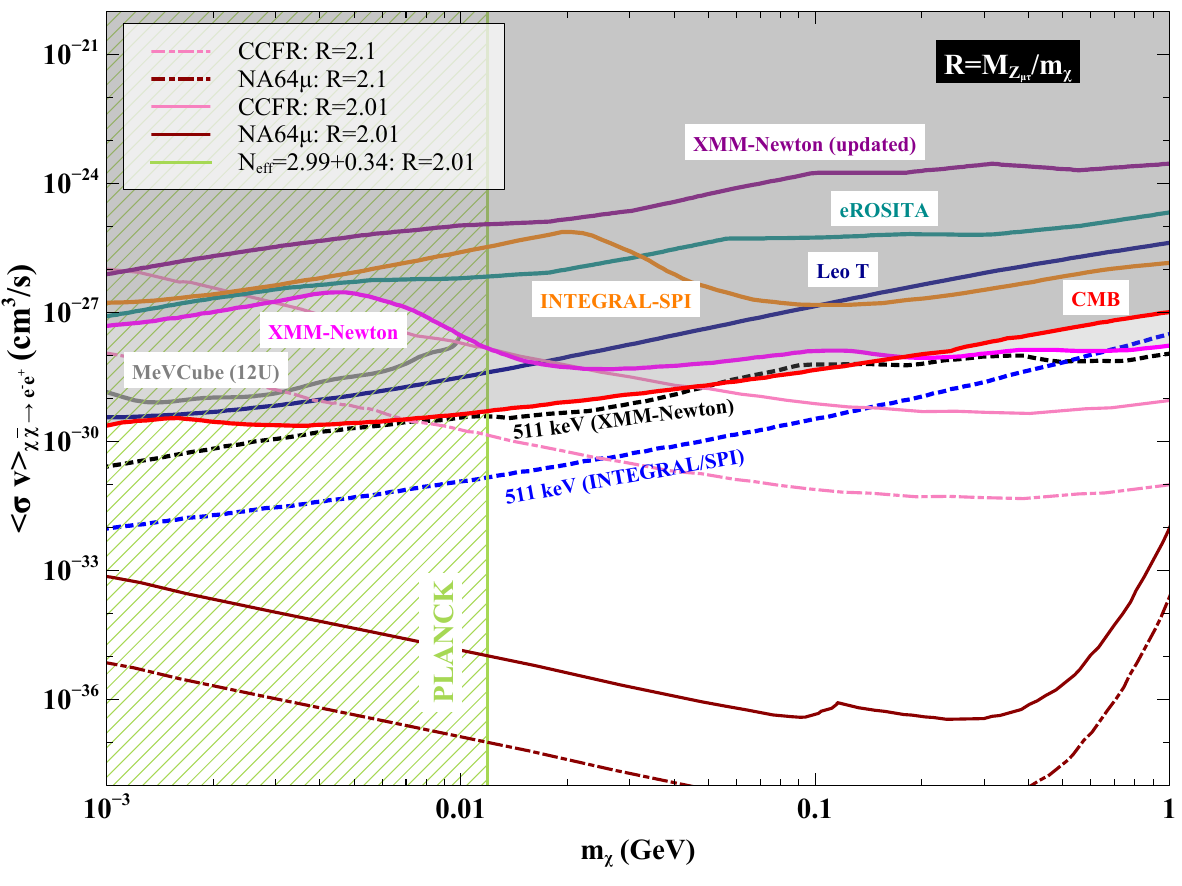}
    \caption{Indirect-detection limits from X-ray observations constrain the annihilation cross-section $\langle \sigma v \rangle_{\chi \bar{\chi} \to e^+ e^-}$. Complementary bounds on the annihilation rate to $e^+ e^-$ are plotted by computing the cross-sections for (R=2.01) and (R=2.1) using constraints from fixed-target experiments, specifically CCFR and NA64$\mu$.}
    \label{fig:inddet}
\end{figure}

Fig.~\ref{fig:inddet} shows the current bounds on $\langle \sigma v\rangle_{\chi\bar{\chi} \to e^+e^-}$, including the CMB limit from Planck \cite{Slatyer:2017sev,ODonnell:2024aaw} (solid red), as well as indirect search constraints from INTEGRAL-SPI \cite{Cirelli:2020bpc} (solid orange), eROSITA \cite{Balaji:2025afr} (dark cyan), Leo T \cite{Wadekar:2021qae} (dark blue) , MeVCube \cite{Saha:2025wgg} (solid grey) and XMM-Newton \cite{Cirelli:2023tnx} (solid magenta). For completeness, we display both the XMM-Newton bound (magenta) from \cite{Cirelli:2023tnx} and the updated bound (dark magenta) reported in \cite{Balaji:2025afr}, where the authors corrected a misapplication of the exposure-weighted solid angle in \cite{Cirelli:2023tnx}. Other probes such as AMS-02~ and Voyager~1 yield significantly weaker limits compared to Planck and are therefore omitted. In addition to X-ray constraints, we incorporate the bound from the 511 keV emmision line \cite{DelaTorreLuque:2023cef,laTorreLuquePedro:2024est} from XMM-Newton (black-dotted) and INTEGRAL/SPI (blue doted), arising from the electron-positron cascade and subsequent positron bound-state formation following DM annihilation to electron-positron pairs. For comparison, we also overlay fixed-target constraints translated from the $(g_{\mu\tau}, M_{Z_{\mu\tau}})$ plane, namely CCFR \cite{Altmannshofer:2014pba} and NA64$\mu$ \cite{NA64:2024klw}. These are shown for representative ratios $R = M_{Z_{\mu\tau}}/m_\chi = 2.01$ (solid) and $R=2.1$ (dot-dashed), providing complementary bounds to the indirect searches. Astrophysical observations, such as white dwarf cooling, also impose constraints \cite{Foldenauer:2024cdp} in the $g_{\mu\tau}-M_{Z_{\mu\tau}}$ parameter space. These arise from the unavoidable loop-induced mixing of the $Z_{\mu\tau}$ boson with photon, which enables its decay into electrons and neutrinos, thereby enhancing the stellar cooling rate. However, this constraint remains comparatively weaker than the bounds presented in Fig.~\ref{fig:summarymutau}. We also check for the Indirect search constraints for $\mu^+\mu^-$ final states which is shown in Appendix~\ref{app::IndirectDmuon}.

% Indirect detection probes astrophysical signatures of DM annihilation or decay into Standard Model particles, such as $\gamma$ rays, neutrinos, or charged cosmic rays, from regions of high DM density (e.g., GC, dSphs, galaxy clusters). In this work we focus on final-state electrons, with a suppressed coupling $g_{\mu\tau}/70$. 
% \begin{figure}[h]
%     \centering
%     \includegraphics[width=0.45\textwidth]{Images/IndirectDetection.png}
%     \caption{Indirect Detection Bound.}
%     \label{fig:A3}
% \end{figure}

% Figure~\ref{fig:A3} shows the current constraints on the annihilation cross section $\sigma_{\chi\chi\to e^+e^-}$, including bounds from Planck \cite{Slatyer:2015jla} (solid red), INTEGRAL-SPI \cite{Cirelli:2020bpc} (solid orange) and XMM-Newton \cite{Cirelli:2023tnx} (solid magenta). Constraints from other experiments such as AMS-02 and Voyager 1 are significantly weaker than the Planck bound and are therefore not shown. 
% For comparison, CCFR \cite{Altmannshofer:2014pba} and NA64$\mu$ bounds \cite{NA64:2024klw}, mapped from the $g_{\mu\tau}–M_{Z_{\mu\tau}}$ plane, are shown for $R=2.1$ (dot-dashed) and $R=3$ (solid), providing complementary constraints.
%We also examined the projected sensitivities from M31 (dashed green) and Draco (both with $10^6$ s exposure) \cite{ODonnell:2024aaw} which likewise remain weaker than the Planck limit. 

\subsubsection*{ Other Constraints and Summary}

\begin{figure}[h]
    \centering
    \includegraphics[width=0.43\textwidth]{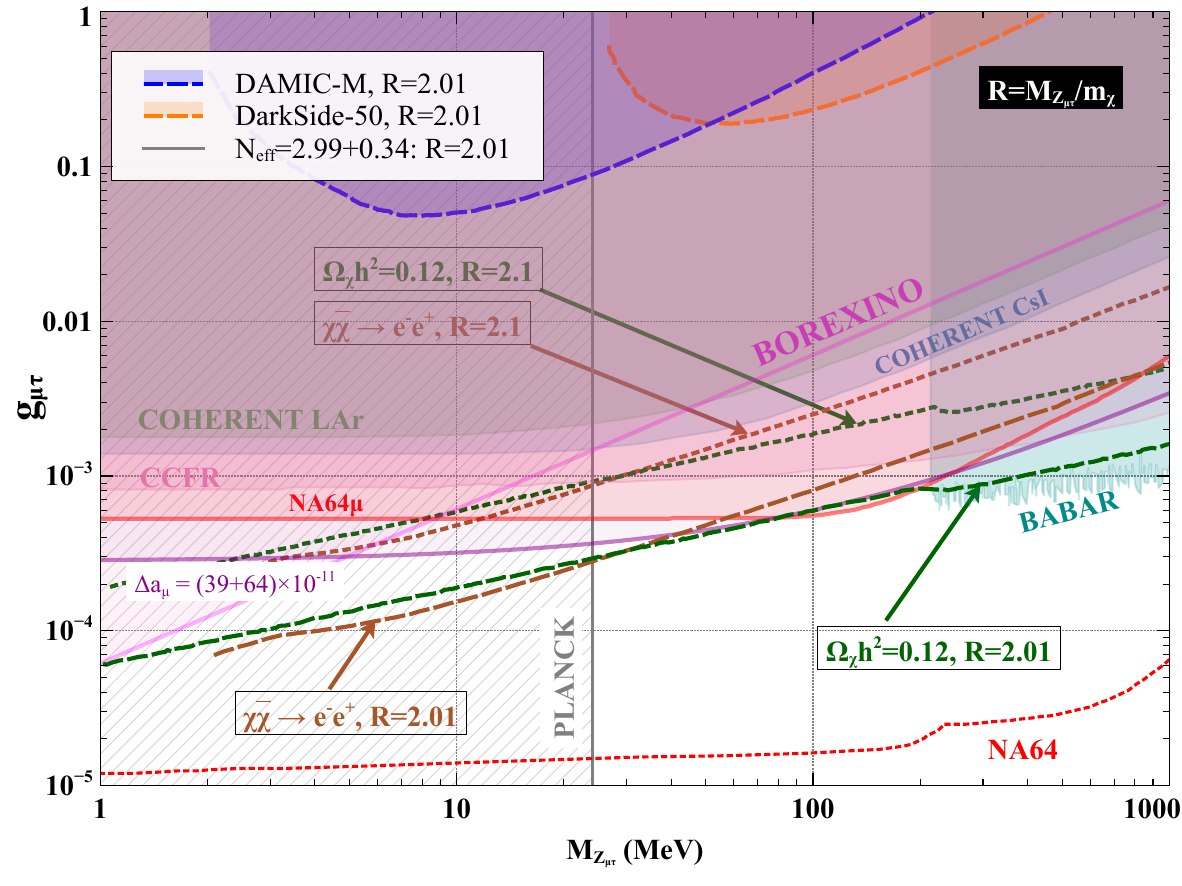}
    \caption{Constraints in the $(g_{\mu\tau},M_{Z_{\mu\tau}})$ plane for two representative choices of the
mass ratio $R \equiv M_{Z_{\mu\tau}}/m_\chi = 2.01$ (dashed lines) and $R=2.1$ (dotted lines).
The blue (DAMIC-M) and orange (DarkSide-50) curves show direct-detection limits from
DM--electron scattering.
Additional exclusions from NA64$\mu$, Borexino, CCFR, COHERENT, BABAR  and $\Delta N_{\rm eff}$ (PLANCK) are overlaid,
along with the CMB bound (brown) shown for both values of $R$
using the corresponding line styles.
The dark-green curves denote the parameter space yielding the observed relic abundance for
$R=2.01$ (dashed) and $R=2.1$ (dotted).
As $R$ is increased away from the resonant value $R=2$, the relic-density contour shifts
to larger values of $g_{\mu\tau}$ due to the reduced resonant enhancement.
The figure highlights that only a narrow window of $M_{Z_{\mu\tau}}$ remains viable,
which will be probed by upcoming NA64 searches.
}
    \label{fig:summarymutau}
\end{figure}

A variety of experimental bounds significantly constrain the parameter space of the $U(1)_{L_\mu-L_\tau}$ model.  In Fig.~\ref{fig:summarymutau}, we show the exclusion limits in the $(g_{\mu\tau}, M_{Z_{\mu\tau}})$ plane for two mass ratio $R=M_{Z_{\mu\tau}}/m_\chi=2.01$ and $2.1$. The most stringent limits from direct-detection arise from DM--electron scattering searches, namely DAMIC-M \cite{DAMIC-M:2025luv} and DarkSide-50 \cite{DarkSide:2022knj}, shown by the blue and orange curves, respectively.  Additional bounds come from accelerator and neutrino experiments, such as NA64$\mu$ \cite{NA64:2024klw}, Borexino \cite{Gninenko:2020xys}, CCFR \cite{Altmannshofer:2014pba}, COHERENT \cite{COHERENT:2017ipa,COHERENT:2020iec}, and BABAR \cite{BaBar:2016sci}, along with the CMB constraint \cite{Slatyer:2015jla} indicated by the brown curve.  For comparison, the upper limit of the anomalous magnetic moment deviation of the muon $\Delta a_\mu$ is also displayed by the purple curve.
Recent results from NA64 and the updated muon $g-2$ measurements already rule out a large fraction of the parameter space, excluding wide regions of gauge coupling $g_{\mu\tau}$ and mediator mass $M_{Z_{\mu\tau}}$.  As expected for $s$-channel annihilation processes, the CMB constraint provides the strongest bound, as it probes the late-time energy injection from DM annihilations at recombination.  

To evade these stringent limits while still accounting for the observed relic abundance, we focus on the resonant annihilation regime, where $m_\chi \simeq M_{Z_{\mu\tau}}/2$.  
This condition is implemented by fixing $R=2.01$, and the resulting relic density contour is shown by the dashed green color in Fig.~\ref{fig:summarymutau}.  
Even under such fine-tuning, we find that viable solutions require $M_{Z_{\mu\tau}} \gtrsim 30$ MeV. At lower mediator masses, the combined CMB and $g-2$ limits entirely exclude the parameter space.  

We now briefly comment on the behavior of the relic-density contour with the variation of the ratio
$R \equiv M_{Z_{\mu\tau}}/m_\chi$.
When $R$ deviates from the resonant value $R=2$, the $s$-channel annihilation
$\chi\bar\chi \to Z_{\mu\tau} \to \mathrm{SM}$ moves away from the resonance and a larger gauge
coupling $g_{\mu\tau}$ is required to reproduce the canonical thermal cross-section.
Hence the relic-density contour shifts to larger $g_{\mu\tau}$ for increasing $|R-2|$.
Conversely, as $R$ approaches 2, the resonant enhancement increases the annihilation
rate and the required coupling $g_{\mu\tau}$ decreases. For example, with the parameters used in Fig.~\ref{fig:summarymutau}, the relic-density contour lies at smaller values of $g_{\mu\tau}$ for $R=2.01$ than for $R=2.1$, and an analogous shift occurs for $R\lesssim2$, reflecting the symmetric behaviour of the resonance around $R=2$.

To give a concrete illustration, consider $m_\chi = 50$ MeV. We find that the relic-density
contour requires
$ g_{\mu\tau} \sim 6.3 \times 10^{-4}$ for $R = 2.01$, 
whereas for
$R = 2.1$ the required coupling increases to 
$g_{\mu\tau} \sim 2 \times10^{-3}$,
pushing the corresponding parameter space into regions already excluded by NA64$_\mu$ and the current $(g-2)_\mu$ upper bound.
This clearly illustrates that departures from the near-resonant regime rapidly reduce the viable parameter
space.

We also account for constraints from $N_{\rm eff}$, arising due to modifications in neutrino decoupling or additional contributions to the neutrino energy density from $\chi\bar{\chi}$ annihilation or mediator decays.  
The details of this analysis are provided in Appendix~\ref{app::Neff}.  
Including these considerations, we find that the gauge boson mass is only viable in a relatively narrow window between $\sim 30$ MeV and a few hundred MeV.  
Beyond this range, the parameter space consistent with DM relic density is again excluded, most notably by the BABAR bound.  Interestingly, the surviving region lies within the projected sensitivity reach of NA64~\cite{Gninenko:2014pea}, implying that the model will be thoroughly tested in the near future.

\section{$\boldsymbol{U(1)_X}$ Model}\label{sec:U1X}
Having discussed the loop-induced kinetic mixing in the $U(1)_{L_\mu-L_\tau}$ case,
let us now consider a more general extension of the SM with an additional Abelian gauge symmetry $U(1)_X$. Such scenarios are theoretically attractive because they allow additional freedom in model building: in particular, the kinetic mixing parameter $\epsilon$ is treated as an independent free parameter, unlike the loop-suppressed mixing in $U(1)_{L_\mu-L_\tau}$. This makes $U(1)_X$ models more flexible in accommodating phenomenological requirements in DM frameworks while retaining a minimal gauge structure.  

A generic $U(1)_X$ introduces a new neutral gauge boson $X_\mu$ (often denoted $Z'$) that can either couple directly to SM fermions through their assigned $U(1)_X$ charges $Q_X^f$ or indirectly via kinetic mixing with the SM hypercharge boson $B_\mu$:
\begin{equation}
	\mathcal{L}_{X} \supset g_d \sum_f Q_X^f \, \bar{f}\gamma^\mu f \, X_\mu \; - \; \frac{\epsilon}{2} B_{\mu\nu}X^{\mu\nu},
\end{equation}
where the second term induces effective couplings between $X_\mu$ and SM fermions even if the latter's $U(1)_X$ charge is zero. We consider this scenario with $Q_X^f=0$ in the remainder of our discussion.

Through this kinetic mixing, $X_\mu$ interacts with the SM fermionic currents as \cite{Borah:2025cqj}
\begin{equation}
	X_\mu j^\mu_{\rm SM} = -X_\mu \, \overline{f}\gamma^\mu \big\{ C^f_L P_L + C^f_R P_R \big\} f ,
	\label{eq:X-interaction}
\end{equation}
where $C^f_L$ and $C^f_R$ denote the effective left- and right-handed couplings of fermion $f$ to $X_\mu$. These couplings can be expressed in terms of the SM gauge couplings as
\begin{equation}
	\begin{split}
		C^f_L &= \left( \sin\eta + \frac{\cos\eta \, \epsilon \, s_w}{\sqrt{1-\epsilon^2}} \right) g^f_L
		- \left( \frac{\cos\eta \, \epsilon \, c_w}{\sqrt{1-\epsilon^2}} \right) e Q^f , \\
		C^f_R &= \left( \sin\eta + \frac{\cos\eta \, \epsilon \, s_w}{\sqrt{1-\epsilon^2}} \right) g^f_R
		- \left( \frac{\cos\eta \, \epsilon \, c_w}{\sqrt{1-\epsilon^2}} \right) e Q^f ,
	\end{split}
\end{equation}
where $s_w = \sin\theta_W$ and $c_w = \cos\theta_W$. The mixing angle $\eta$ quantifies the mass mixing between the $X$ boson and the $Z$ boson, given approximately by
\begin{equation}
	\tan\eta \simeq \frac{2 \epsilon s_w}{1 - \frac{M_X^2}{M_Z^2}} .
\end{equation}
For reference, the usual SM couplings to $Z$ boson are
\begin{equation}
	\begin{split}
		g_L^f &= \frac{g}{c_w}\left(T^f_{3L} - s_w^2 Q^f\right), \qquad
		g_R^f = -\frac{g}{c_w} \, s_w^2 Q^f ,
	\end{split}
\end{equation}
where $T^f_{3L}$ and $Q^f$ are the weak isospin and electric charge of fermion $f$, respectively. 

The mass of the new gauge boson $X_\mu$ can originate either from the Stueckelberg mechanism \cite{Stueckelberg:1938hvi} or from the Higgs mechanism involving a scalar field charged under the new $U(1)_X$, depending on the specifics of the model construction.

\subsubsection*{Dark matter self-interactions}\label{subs:SIDM}

Self-interacting dark matter (SIDM) has gained significant attention as a compelling solution to small-scale structure issues in astrophysics. These observations favor dark matter models with sizable self-interaction cross-sections, typically $\sigma/m_\chi \sim 0.1 - 10 \, \mathrm{cm}^2/\mathrm{g}$ which can thermalize dark matter within halos and alleviate tensions with collisionless cold dark matter predictions~\cite{Spergel:1999mh, Tulin:2017ara, Bullock:2017xww}.

Unlike the $U(1)_{L_\mu-L_\tau}$ model where gauge $g_{\mu \tau}-M_{Z_{\mu \tau}}$ plane is tightly constrained in the sub-GeV regime, a dark $U(1)_X$ allows the possibility of a light gauge boson $X_\mu$ together with a relatively strong coupling $g_d$ which can generate the required DM self-interactions via light gauge boson $X_\mu$ exchange. Interestingly, DM self-interactions arising via light mediator exchange are velocity-dependent thereby solving the small-scale issues at dwarf galaxy scales while being consistent with standard CDM properties at large or cluster scales \cite{Buckley:2009in, Feng:2009hw, Feng:2009mn, Loeb:2010gj, Bringmann:2016din, Kaplinghat:2015aga, Aarssen:2012fx, Tulin:2013teo}.

The relevant DM Lagrangian is $\mathcal{L} \supset - g_d \, \overline{\chi} \gamma^\mu \chi X_\mu$, describing dark matter self-interactions mediated by gauge boson $X$. At non-relativistic velocities, the effective potential for dark matter self-scattering is given by the Yukawa form
\begin{equation}
	V(r) = \pm \frac{\alpha_d}{r} e^{-M_X r},
\end{equation}
where $\alpha_d = g_d^2/(4\pi)$, and the $-$ sign corresponds to attractive interaction for opposite charges (particle-antiparticle) and the $+$ sign for repulsive interaction for same charges (particle-particle). The detailed expressions for the self-interaction cross-sections are provided in Appendix~\ref{app::sidm_cross-section}. 

It should be emphasized that, although the $U(1)_X$ setup offers more freedom compared to the $U(1)_{L_\mu-L_\tau}$ case due to the additional kinetic mixing parameter, the requirement of sizable self-interactions, correct relic density of dark matter and other constraints imposes strong bounds on the allowed $(g_d, m_\chi, M_X)$ parameter space. As a result, the viable region remains highly constrained and hence predictive.

% As the model contains one additional free parameter $\epsilon$, the direct-detection cross-section (Eqn.\ref{eq:DD_U1D}) implies that smallness of $\epsilon$ leads a naturally larger $g_d$, which can be brought in the desired ballpark of dark matter self-interactions.

% In this $U(1)_X$, the relevant Lagrangian is $\mathcal{L}\supset -g_d \overline{\chi}\gamma^\mu \chi X_\mu$. The DM interacts with themselves ($\chi \chi \leftrightarrow \chi \chi$) via the exchange of the $X$ boson. The non-relativistic effective potential for the above interaction is given by
% \begin{equation}
%     V(r)= \pm \frac{\alpha_d}{r}~e^{-r/M_X^{-1}} 
% \end{equation}
% where $\alpha_d=g_d^2/4\pi$. The cross-sections for self-interactions is listed in Appendix \ref{app::sidm_cross-section}

\begin{figure}[h]
    \centering
    \includegraphics[width=0.43\textwidth]{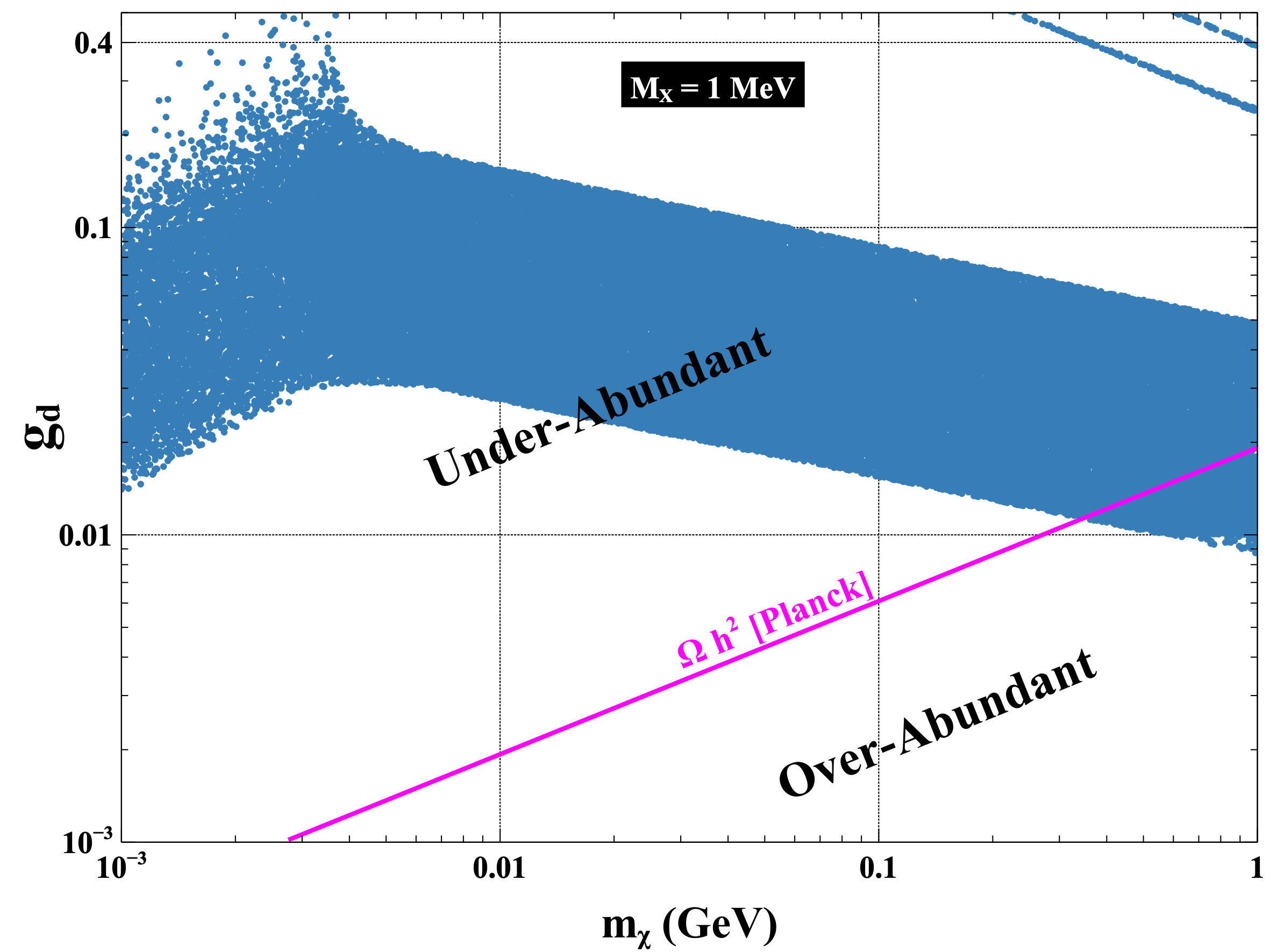}
    \caption{Parameter space in the $g_d$–$m_\chi$ plane illustrating the region consistent with self-interaction criteria for $M_X = 1 \,\text{MeV}$, together with the correct relic density constraint (magenta line) in the minimal setup. Parameter space above the magenta line give an under-abundant relic density, while points below the line are over-abundant. The large blue region indicates the Born-regime of self-interactions while the thin lines in the upper-right corner depicts the transition to the resonant regime (see
Appendix~\ref{app::sidm_cross-section}).}
    \label{fig:B2}
\end{figure}

\subsubsection*{Relic Density}
In this setup, the relic abundance of dark matter is primarily determined by the annihilation process $\chi \bar{\chi} \to X^\mu X_\mu$ as opposed to annihilation into SM fermions, since the latter channel is relatively suppressed due to small kinetic mixing~\cite{Borah:2022ask,Borah:2023sal,Mahapatra:2023oyh,Borah:2024wos,Adhikary:2024btd}. This leads to the crucial observation that the relic density calculation is, to leading order, independent of $\epsilon$ which instead governs the detection prospects through interactions with visible matter. Even if the small kinetic mixing is insufficient to keep the dark and visible sectors in equilibrium, there exists other portals to do so. For example, in scenarios where the $U(1)_X$ symmetry is broken spontaneously by a scalar $S$, thermal contact between the dark and visible sectors can still be maintained via the quartic portal interaction $(S^\dagger S)(H^\dagger H)$ with the SM Higgs. By assigning $S$ an appropriate non-trivial $U(1)_X$ charge, its direct Yukawa coupling to DM can be forbidden, thereby leaving the phenomenology discussed here unaffected while ensuring thermal equilibrium at early times.

The thermally averaged annihilation cross-section for dark matter into mediator pairs is \(s\)-wave dominated and given by
\begin{equation}
	\langle \sigma v \rangle_{\chi\chi\rightarrow XX} = \frac{g_d^4}{16 \pi m_\chi^2} \sqrt{1 - \frac{M_X^2}{m_\chi^2}}.
    \label{eq:SIDM_th_avg}
\end{equation}
As a sizable coupling \(g_d\) is required to achieve the correct DM self-interactions, this simultaneously enhances the annihilation rate, reducing the relic abundance to the observed range, even for GeV or sub-GeV scale dark matter. This behavior is illustrated in Fig.~\ref{fig:B2}, where the contour of correct relic density is displayed by the solid magenta colored line along with the green shaded region of parameter space consistent with the required DM self-interaction.
%The blue and purple shaded regions represent the SIDM parameter space for two fixed gauge boson masses $M_X=1~{\rm MeV}$ and $M_X=10~{\rm MeV}$, respectively. 

% The same Lagrangian (subsection ~\ref{subs:SIDM}) also dictates the possibility of establishing the relic density of dark matter via $\chi \chi \leftrightarrow X^\mu X_\mu$. This annihilation is s-wave dominated and the thermal average is conveniently expressed by
% \begin{equation}
%     \langle \sigma v \rangle_{\chi\chi\rightarrow XX} = \frac{g_d^4}{16 \pi m_\chi^2}~\sqrt{1-\frac{M_X^2}{m_\chi^2}}
% \end{equation}

% Since a large $g_d$ is required to account for the self-interactions of dark matter, it also leads to enhanced annihilation rates, which in turn reduce the relic density of SIDM across much of the parameter space, as illustrated in Fig.~\ref{fig:S1}.

\subsubsection*{Direct detection}
Direct detection of dark matter in this framework relies on the small but universal coupling of the $X$ boson to the SM fermions through kinetic mixing. For DM in the $1~\mathrm{MeV}$–$1~\mathrm{GeV}$ mass range, the dominant DM-electron scattering arises due to exchange of $X$ boson.

In the limit of very small $\epsilon$, the induced vector coupling of the $X$ boson to electrons is
\begin{equation}
	g_V^e = \frac{C_L^e+C_R^e}{2} \;\simeq\; 0.26\, \epsilon ,
\end{equation}
which governs the DM–electron scattering process. The corresponding cross-section takes the form
\begin{equation}
	\sigma (\chi e \leftrightarrow \chi e)
	= \frac{\mu_{\chi e}^2}{\pi} \;
	\frac{(g_V^e)^2 g_d^2}{\big(M_X^2+\alpha^2 m_e^2\big)^2},
	\label{eq:DD_U1D}
\end{equation}
where $\mu_{\chi e}$ is the reduced mass of the DM–electron system.

% Dark matter–electron scattering arises from the universal coupling of the $X$ boson to SM fermions through kinetic mixing, which in particular induces interactions with electrons.
% For dark matter in the $1\ \mathrm{MeV} $–$1\ \mathrm{GeV}$ range, these scattering constraints remain relevant.
% In the limit of very small $\epsilon$, the vector coupling governing $\chi$–$e$ scattering is given by

% \begin{equation}
%     g_V^e = \frac{C_L^e+C_R^e}{2}~\simeq 0.26 \epsilon
% \end{equation}
% Then the relevant cross-section for dark matter-electron is shown as
% \begin{equation}
%     \sigma (\chi e \leftrightarrow \chi e)=\frac{\mu_{\chi e}^2}{\pi}~\frac{(g_V^e)^2 g_{d}^2}{(M_X^2+\alpha^2 m_e^2)^2}
%     \label{eq:DD_U1D}
% \end{equation}

\begin{figure}[h]
    \centering
    \includegraphics[width=0.43\textwidth]{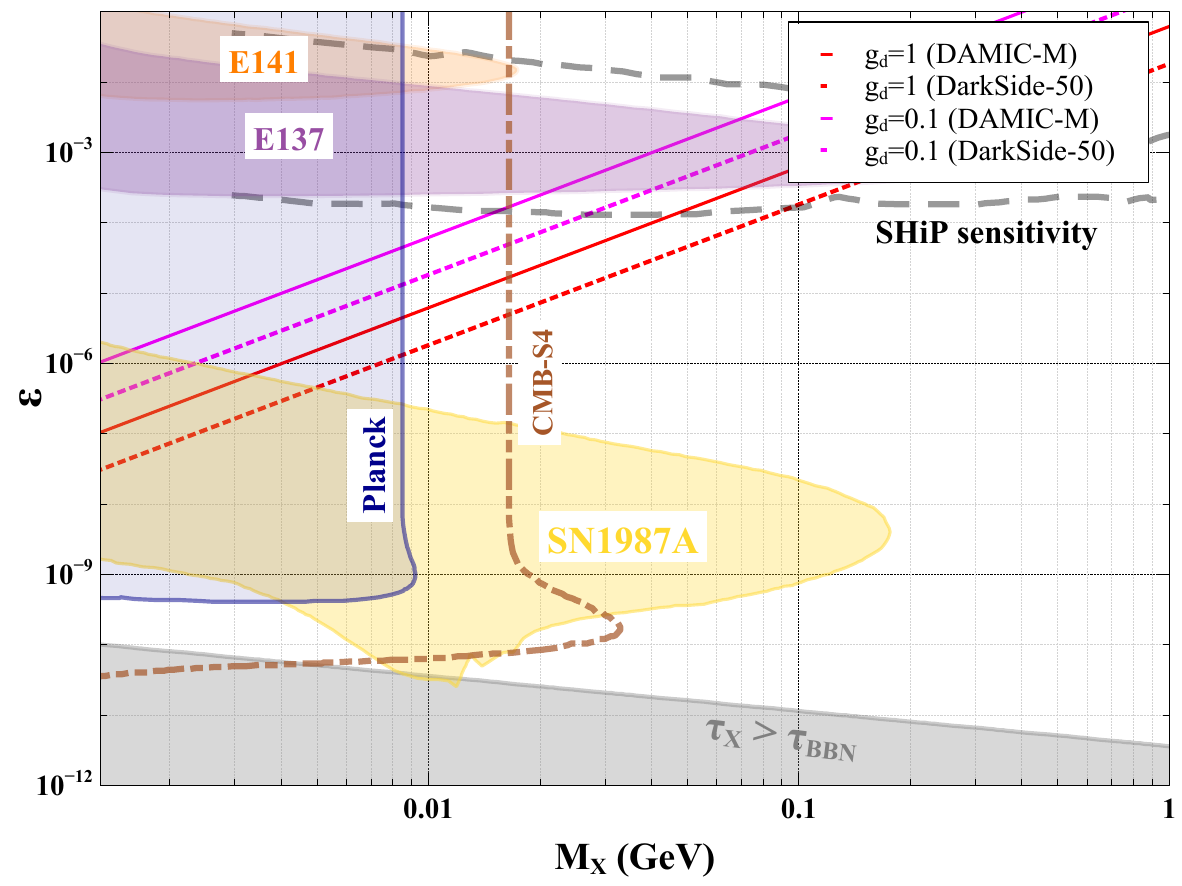}
    \caption{Parameter space in $\epsilon-M_X$ plane of the $U(1)_X$ model after imposing various constraints. The most stringent bounds from DAMIC-M ($m_\chi$=10 MeV) and DarkSide-50 ($m_\chi$=100 MeV) for two different values of gauge coupling $g_d$= are shown by solid and dashed lines, respectively. Other shaded regions and contours correspond to different bounds or future sensitivities, see main text for details.
    }
    \label{fig:B1}
\end{figure}  
Since the direct-detection cross-section in Eq.~\eqref{eq:DD_U1D} depends on the product $\epsilon^2 g_d^2$, the limits from electron recoil searches can be reinterpreted in the $\epsilon$–$M_X$ parameter space for given values of $g_d$. To determine the allowed value of $\epsilon_{DD}$ for a given $g_d$, we use the most sensitive cross-sections from DAMIC-M and DarkSide-50, corresponding to dark matter masses of $m_\chi \sim 10$ MeV and $m_\chi \sim 100$ MeV, respectively. Since for these masses the reduced mass is approximately equal to the electron mass, we obtain the following:
\begin{equation}
    \epsilon_{DD} = 
\begin{cases}
  5.508 \times 10^{-5} \left(\frac{0.1}{g_d}\right) \left(\frac{M_X({\rm GeV})}{0.01}\right)^2:~ \text{DAMIC-M} \\
  1.818 \times 10^{-5} \left(\frac{0.1}{g_d}\right) \left(\frac{M_X({\rm GeV})}{0.01}\right)^2:~ \text{DarkSide-50}
\end{cases}
\label{eq:epsilon_DD}
\end{equation}
Eq.~\ref{eq:epsilon_DD} gives the most stringent upper bound on the kinetic mixing parameter $\epsilon$ from the direct detection experiments DAMIC-M and DarkSide-50. For other dark matter masses, the allowed values of $\epsilon$ are larger, with an explicit dependence on $m_\chi$ appearing near $m_e$. This behavior is illustrated in Fig.~\ref{fig:B1}, where we show the strongest constraints from DAMIC-M ($m_\chi = 10$ MeV) and DarkSide-50 ($m_\chi = 100$ MeV) for two representative couplings, $g_d = 0.1$ and $g_d = 1$. 
The same constraints can be displayed in the $\sigma_e$ vs $m_\chi$ plane for fixed values of $M_X$ and $\epsilon$, highlighting the regions consistent with relic density and self-interaction requirements. The corresponding figure is provided in Appendix~\ref{ddu1x}.
%The same constraints can also be shown in the $g_d$ vs $m_\chi$ plane for fixed values of $M_X$ and $\epsilon$ along with the viable parameter space to achieve the correct relic density and the required self-interaction.  This can be seen in Appendix~\ref{ddu1x}.  

\subsubsection*{Summary of Other Constraints}
Fig.~\ref{fig:B1} also displays existing experimental bounds related to several observables. The purple and orange shaded regions are excluded by electron beam-dump experiments such as SLAC-E141, SLAC-E137 \cite{Bjorken:2009mm, Andreas:2012mt}, and Fermilab-E774 \cite{PhysRevLett.67.2942}, where dark gauge bosons are produced via Bremsstrahlung processes and subsequently escape detection. The projected sensitivity of the SHiP experiment at CERN \cite{SHiP:2015vad}, designed to probe hidden sectors in conjunction with tau neutrino measurements, is also illustrated. Additionally, the yellow shaded region shows constraints from Supernova 1987A \cite{Chang:2016ntp}, which arise from the requirement that excessive dark gauge boson production should not suppress the observed neutrino flux. Constraints from $\Delta N_{\rm eff}$ limit the viable parameter space as well. For mediator masses below 10 MeV, only very small mixings ($\epsilon \lesssim 10^{-9}$) are allowed, as can be seen from the white patch currently allowed by PLANCK, SN1987A and lifetime bounds. On the other hand, for heavier mediators ($M_X > 10$ MeV) the limit weakens to $\epsilon \lesssim 10^{-4}$, as illustrated by the DAMIC-M (solid) and DarkSide-50 (dotted) curves for two specific choices of $g_d=0.1$ (magenta) and $g_d=1$ (red). The dot-dashed brown line denotes the sensitivity of the proposed CMB-S4 experiment \cite{CMB-S4:2016ple}. Finally, the Big Bang Nucleosynthesis (BBN) bound is also shown, derived by demanding the lifetime of the dark gauge boson $X$ to be shorter than typical BBN timescale.

It is worth noting that in this setup the mediator mass $M_X$ is ideally required to be at or below the MeV scale or the electron mass threshold, ensuring that it decays exclusively into SM neutrinos. For heavier mediators, one must instead guarantee that $X$ decays dominantly into some form of dark radiation rather than SM charged states. This choice is crucial to evade stringent indirect-detection limits, as well as constraints from CMB measurements on dark matter annihilation into charged final states mediated by $X$ bosons~\cite{Madhavacheril:2013cna, Slatyer:2015jla, Planck:2018vyg, Elor:2015bho, Profumo:2017obk} of the type $\chi \bar{\chi} \to X X \to 4\,{\rm SM}$.

% Figure~\ref{fig:B2} presents the parameter space relevant for dark matter self-interactions in the $g_d$–$m_\chi$ plane, shown for two benchmark values of $M_X$: 10 MeV (blue points) and 100 MeV (purple points). The red dashed line indicates the direct detection constraint from DAMIC and DarkSide-50, specifically for $M_X = 10$ MeV and $\epsilon = 10^{-5}$. This illustrates that by treating $\epsilon$ as a free parameter, certain regions of the parameter space can evade bounds from dark matter–electron scattering.

% Notably, for $\epsilon < 10^{-9}$, all the points corresponding to $M_X = 10$ MeV (blue) remain safe from direct detection constraints. However, these same points fail to produce the observed dark matter relic abundance within the minimal model. The correct relic density can only be achieved for the points that lie along the orange line, which represents the parameter combinations consistent with thermal freeze-out.

\section{Conclusion}\label{sec:conclusion}
In this work, we have analyzed two well-motivated gauge extensions of the Standard Model namely, the gauged $U(1)_{L_\mu - L_\tau}$ and a dark $U(1)_X$ scenario in the context of sub-GeV dark matter which is facing tight constraints form direct-detection experiments like DAMIC-M probing DM-electron scattering. Both the frameworks possess distinctive theoretical and phenomenological appeals and offer viable mechanisms to explain the observed dark matter relic abundance while connecting to broader open questions such as lepton flavor structure, the muon anomalous magnetic moment, and small-scale structure problems of CDM. We have scrutinized these models in light of the latest and significantly improved electron recoil constraints from the DAMIC-M experiment, which severely restrict dark matter–electron scattering cross-sections. Our comprehensive analysis demonstrates that viable regions of parameter space remain in both models which can reproduce the correct relic density while satisfying stringent bounds from direct detection, accelerator searches, cosmology, and astrophysical observations.

The $U(1)_{L_\mu - L_\tau}$ scenario, motivated by its theoretical minimality, anomaly cancellation, and direct connection to the muon $(g-2)$ anomaly, faces increasingly tight constraints from CMB observations, muon anomalous magnetic moment measurements, and accelerator experiments such as NA64. Despite these challenges, a narrow window of parameter space survives due to resonantly enhanced annihilation near $m_\chi \simeq M_{Z_{\mu\tau}}/2$. This near-resonance condition not only permits the observed relic abundance but also keeps the model testable in upcoming experiments, especially NA64 and future low-threshold direct-detection efforts.

In contrast, the generic $U(1)_X$ extension provides additional model-building freedom by treating the kinetic mixing parameter as a free input rather than a loop-induced quantity. A notable and attractive feature of this framework is the decoupling of the process responsible for generating thermal relic from the one determining DM-electron scattering rates. This occurs due to the possibility of dominant DM annihilation into light dark gauge bosons rather than SM fermions, rendering the relic abundance largely independent of the kinetic mixing parameter. Moreover, the presence of a light mediator naturally facilitates velocity-dependent dark matter self-interactions, offering a compelling resolution to small-scale structure problems that remain unexplained within standard cold dark matter paradigms. Imposing the dual requirements of viable self-interactions and consistency with CMB constraints restricts the viable dark matter mass range to roughly the 300 MeV–1 GeV ballpark, whereas relaxing the self-interaction requirement expands the parameter space considerably.

To conclude, this study points out two interesting ways of realizing sub-GeV thermal DM with a new Abelian gauge interaction which can evade the stringent DAMIC-M constraints on DM-electron scattering. One possibility is to consider family non-universal $U(1)$ gauge charge with resonantly enhanced DM annihilation into SM. Another appealing way is to consider a dark $U(1)$ gauge symmetry with family universal couplings via kinetic mixing where DM dominantly annihilates into light $U(1)$ gauge boson pairs. We considered gauged $L_\mu-L_\tau$ as an illustrative example for the first class of models and compare the sub-GeV DM parameter space with constraints form DAMIC-M, recently updated muon $(g-2)$ world average as well as other bounds from particle physics, cosmology and astrophysics related observations. The second class of models, on the other hand, comes with additional advantage of generating large DM self-interactions pitentially solving the small-scale issues of cold DM. A major part of the currently allowed parameter space of both the models remain within reach of future experiments keeping them testable in near future.

\acknowledgments
We are grateful to Shyam Balaji for bringing to our attention the updated experimental bound from XMM-Newton and eROSITA, which has been incorporated in this work. The work of D.B. is supported in part by the Science and Engineering Research Board (SERB), Government of India grant MTR/2022/000575 and the Fulbright-Nehru Academic and Professional Excellence Award 2024-25. We acknowledge helpful conversations with colleagues during the International Conference - Phoenix (2025), which contributed to shaping parts of this study.

 \section*{}
 \appendix\label{appendix}

\section{Decay Width of a Gauge Boson}\label{app::decay_width}
The decay width of a gauge boson, say $X$ of $U(1)_X$, to a charged lepton with mass $m_f$ and neutrino, with the corresponding couplings $C_L^f,C_R^f$ and $C_L^\nu$ are calculated as 
\begin{equation}
    \Gamma_{X \rightarrow \nu_i\overline{\nu_i}}= \frac{(C^{\nu_i}_L)^2 M_X}{24 \pi},
\end{equation}
\begin{equation}
    \Gamma_{X \rightarrow f\overline{f}}= \frac{(C^f_{LR})^2 M_X}{24 \pi} \sqrt{1-\frac{4 m_f^2}{M_X^2}},
\end{equation}
where the factor $(C^f_{LR})^2$ is given by
\begin{equation}
\left( (C^f_L)^2+(C^f_R)^2 \right) \left(1-\frac{m_f^2}{M_X^2}\right) + 6 \frac{m_f^2}{M_X^2} C^f_L C^f_R.
\end{equation}
In the $U(1)_{L_\mu-L_\tau}$ model, the gauge boson $Z_{\mu\tau}$ has a mass $M_{Z_{\mu\tau}}$, and its couplings to both left- and right-handed fermions are identical, i.e. $C_L^{i}=C_R^{i}=C_L^{\nu_i}=g_{\mu\tau}$ where $i=\{\mu,\tau\}$, while for electron, $C_L^e=C_R^e=g_{\mu\tau}/70$.

\section{Indirect Detection}\label{app::IndirectDmuon}
% Indirect searches for dark matter (DM) candidates rely on astrophysical and cosmological observations — for example, data from Planck, XMM-Newton, etc.— by looking for signatures of DM annihilation into SM particles. In the $U(1)_{L_\mu-L_\tau}$ model, if the DM mass lies below the muon threshold, the only available annihilation channels at tree level are into neutrinos via the gauge coupling $g_{\mu\tau}$. A loop-induced kinetic mixing generates an effective coupling of order $g_X/70$ to electrons, opening a suppressed but phenomenologically important annihilation channel to $e^+e^-$. This channel is relevant because indirect searches are generally more sensitive to charged-lepton final states than to neutral ones.

\begin{figure}[h]
    \centering
    \includegraphics[width=0.43\textwidth]{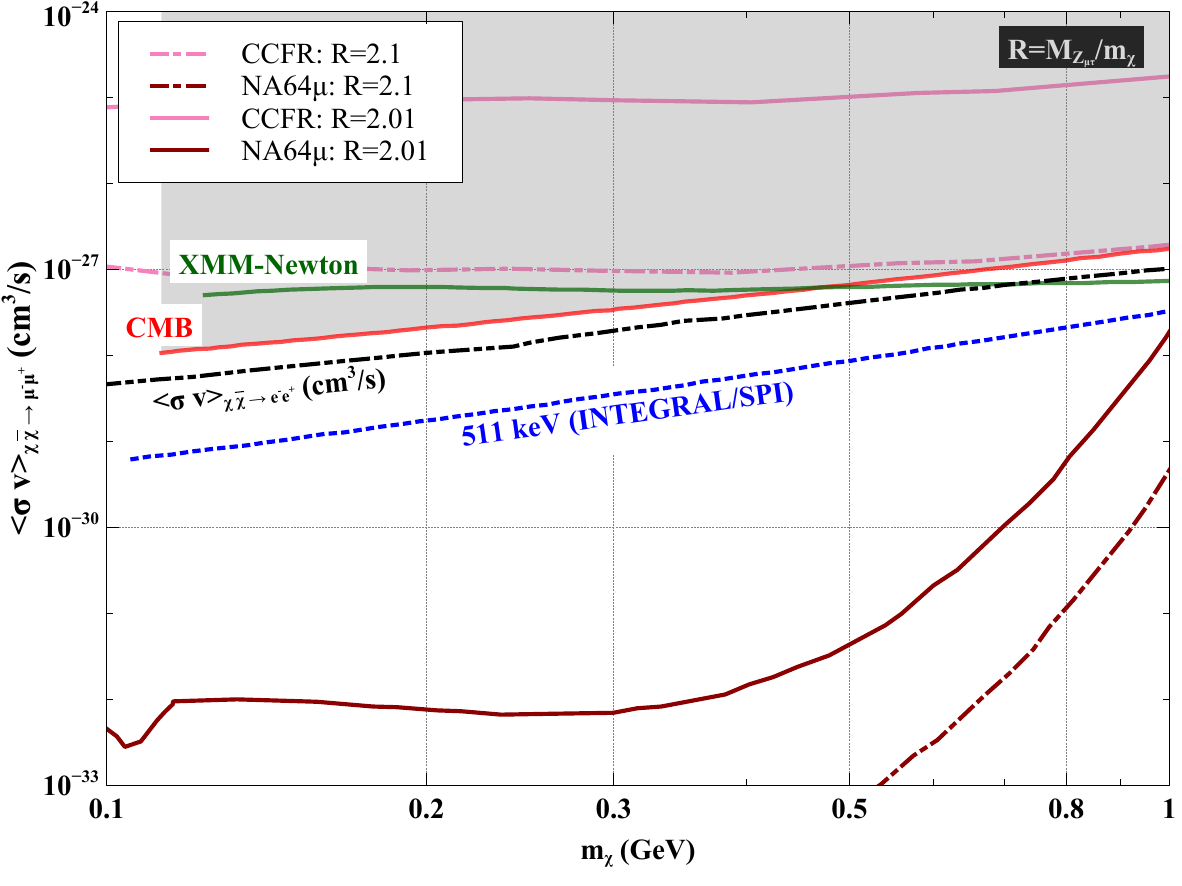}
    \caption{Indirect detection bounds from DM annihilation into $\mu^+\mu^-$. For DM masses above the muon threshold, the constraint from the $\chi \overline{\chi} \rightarrow e^+e^-$ channel (magenta line) is significantly stronger than that from the $\mu^+\mu^-$ final state (red line).}
    \label{fig:inddet_muon}
\end{figure}

In the $U(1)_{L_\mu-L_\tau}$ model, for DM masses above the muon threshold, the model permits direct annihilation into $\mu^+\mu^-$. This channel introduces additional constraints from CMB, though they are somewhat weaker compared to those from the electron channel \cite{Slatyer:2017sev,ODonnell:2024aaw} (dotted magenta line in Fig.~\ref{fig:inddet_muon}). Fig.~\ref{fig:inddet_muon} presents the corresponding exclusion limits from {Planck} \cite{ODonnell:2024aaw} (red shaded) and {XMM-Newton} \cite{Cirelli:2023tnx}(green shaded), together with the annihilation into $e^+e^-$ (dotted-magenta line) for direct comparison. 
%While Fig.~\ref{fig:inddet_muon} illustrates the parameter space for DM annihilation into muon final states, we have also included an exclusion curve corresponding to annihilation into electrons. 
In the $U(1)_{L_{\mu}-L_\tau}$ model, DM annihilation into $e^-e^+$ channel is suppressed by factor of $\epsilon_A^2$ , in comparison to $\mu^-\mu^+$ channel. Therefore, $\chi \overline{\chi} \rightarrow \mu^+\mu^-$ gives a stronger constraint on $g_{\mu\tau}$ in comparison to $\chi \overline{\chi} \rightarrow e^+e^-$. In other words, the muon final state channel will rule out a larger parameter space on $g_{\mu\tau}$ in comparison to electron final state in Fig~\ref{fig:summarymutau}. The complementary sensitivities projected from fixed-target experiments are also shown, depicted by purple (NA64$\mu$) and blue (CCFR) curves for $R=2.1$ (dot-dashed) and $R=2.01$ (solid).

\section{$N_{\rm eff}$ constraints}\label{app::Neff}
If DM mass in the $U(1)_{L_\mu - L_\tau}$ model is below muon mass threshold $m_\mu$, its annihilation proceeds mainly into neutrinos, allowing us to treat $\chi$ and $\nu$ as being in thermal equilibrium. For temperatures below the neutrino decoupling temperature, $T_D = 2.3$ MeV, energy injected from DM annihilation increases the total energy density of the neutrino bath. This increase is quantified by $N_{\rm eff}$, which has a SM value of 3.045 \cite{Mangano:2005cc, Grohs:2015tfy,deSalas:2016ztq}. The additional contribution from $X$-boson decay and DM annihilation can be obtained via entropy conservation and is given by \cite{Boehm:2013jpa}

\begin{equation}
    N_{\rm eff} = N_\nu \left[ 1+\frac{1}{N_\nu} \sum_{i=\chi,Z_{\mu\tau}} \frac{g_i}{2} F\left(\frac{m_i}{T_D}\right) \right]^{4/3}
\end{equation} 
where the function $F(x)$ is given by

\begin{equation}
    F(x)= \frac{30}{7 \pi^4} \int_x^{\infty} dy \frac{(4y^2-x^2)\sqrt{y^2-x^2}}{e^y \pm 1}
\end{equation}
where the +(-) sign refers to fermion(boson) statistics.

\begin{figure}[h]
    \centering
    \includegraphics[width=0.43\textwidth]{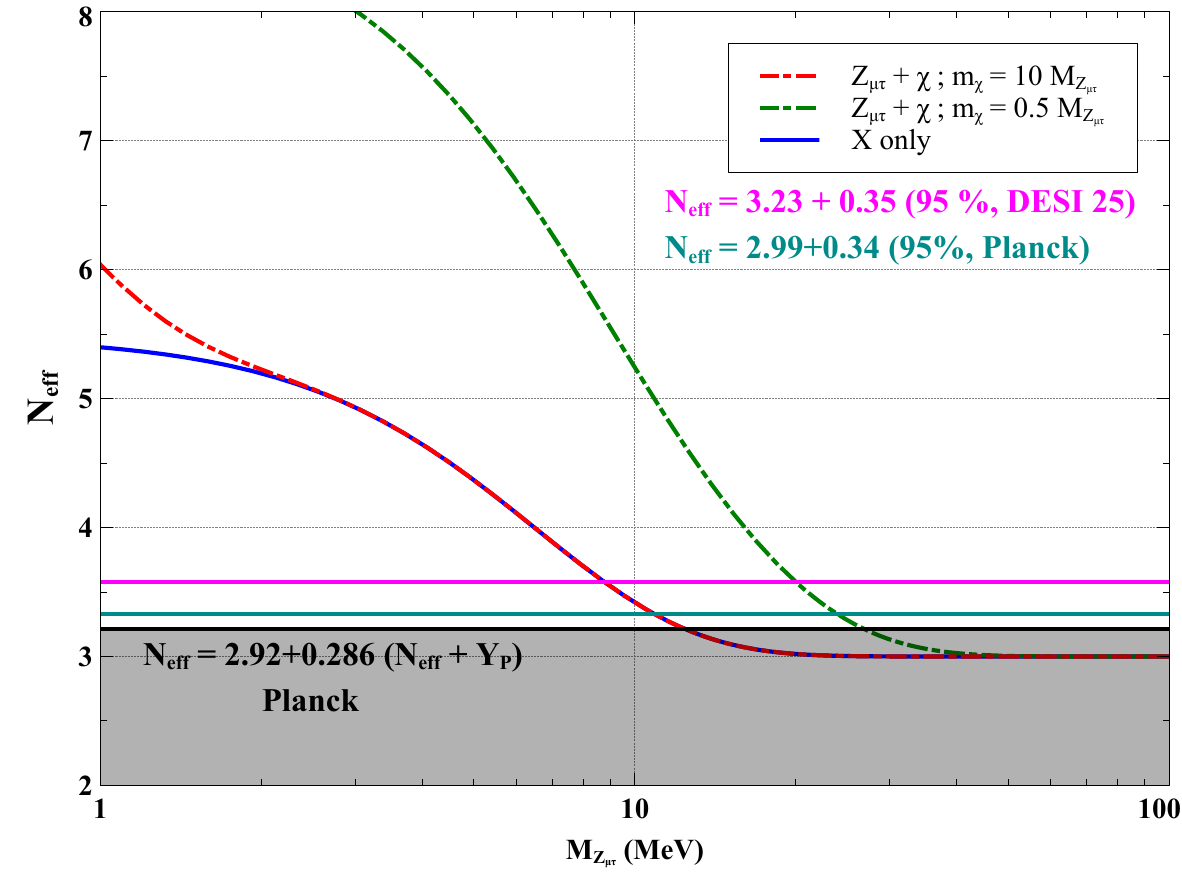}
    \caption{Contribution to the $N_{\rm eff}$, arising purely from the $Z_{\mu\tau}$ boson decay (blue), and additional contribution from $\chi$ annihilation for two different values of $m_\chi$ (dot-dashed red and green lines).}
    \label{fig:APP1}
\end{figure}

Fig.~\ref{fig:APP1} depicts the contribution to $N_{\rm eff}$ arising from the decay process $X \to \nu$, together with the additional contribution from DM annihilation. In the regime $m_\chi > M_{Z_{\mu\tau}}$, the annihilation contribution is negligible, whereas it becomes relevant for $m_\chi < M_{Z_{\mu\tau}}$. The figure also reports the current upper bounds on $N_{\rm eff}$. Specifically, the $95\%$ C.L. constraint from Planck+lensing+BAO \cite{Planck:2018vyg} is given by
$$N_{\rm eff}=2.99^{+0.34}_{-0.33} \quad (95\%),$$
while a fit including the six $\Lambda$CDM parameters extended by $N_{\rm eff}$ and $Y_{\rm P}$ \cite{Planck:2018vyg,2018_Planck_Supp}, as presented in \cite{Giovanetti:2021izc}, yields
$$N_{\rm eff} = 2.926 \pm 0.286 \quad (95 \%).$$
Furthermore, the combination of DESI BAO with CMB data \cite{DESI:2025ejh} within the one-parameter extension $\Lambda$CDM+$N_{\rm eff}$ provides the constraint
$$N_{\rm eff}=3.23^{+0.35}_{-0.34} \quad (95\%).$$
Taken together, these bounds further tighten the parameter space and shift the lower limit on the DM mass inferred from $N_{\rm eff}$ to smaller values.

\begin{figure}[h]
    \centering
    \includegraphics[width=0.43\textwidth]{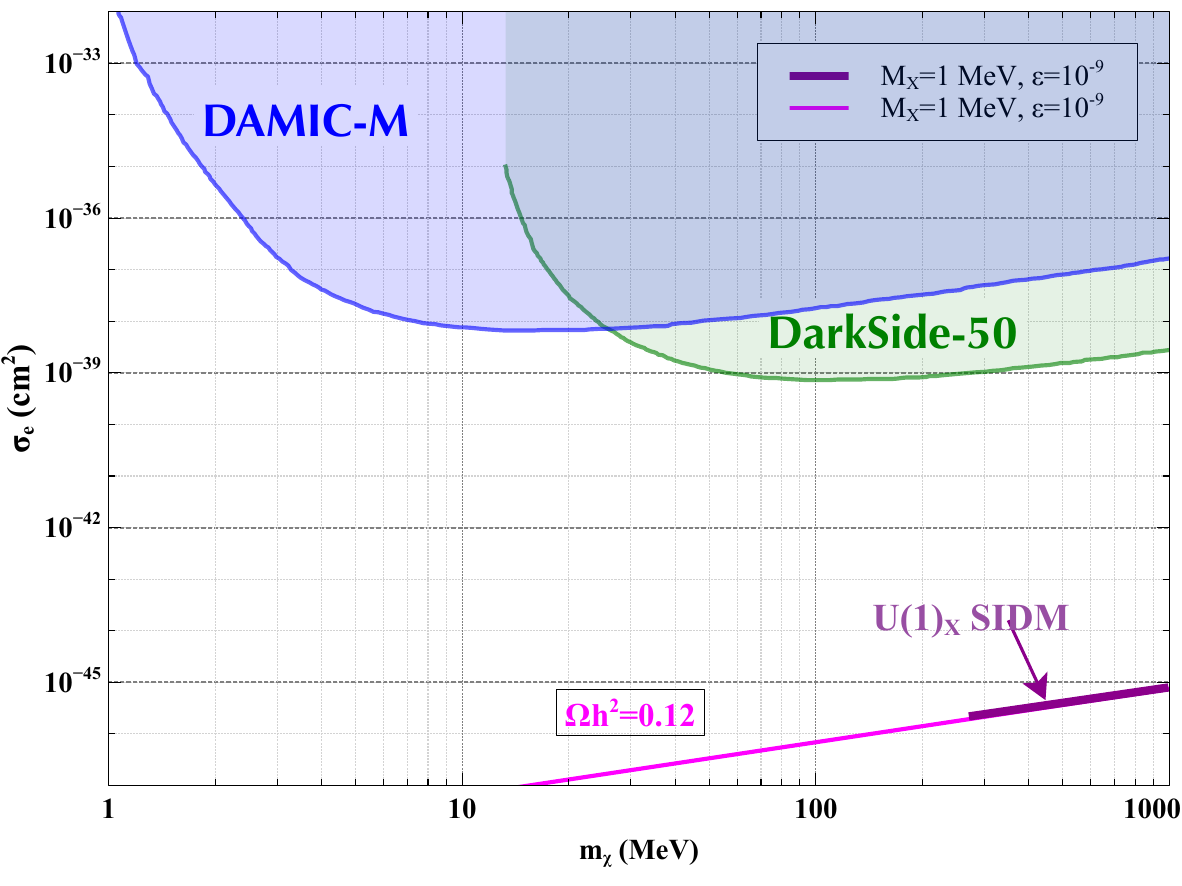}
    \caption{Parameter space for the correct DM relic density, shown in the $\sigma_e$–$m_\chi$ plane, compared with current bounds from DAMIC-M and DarkSide-50. The magenta line indicates the region consistent with the relic density requirement alone, while the purple line highlights the subset of parameter space that simultaneously satisfies both the relic density and self-interaction constraints.}
    \label{fig:S1}
\end{figure}

\section{Low energy cross-sections relevant for the self-interactions of dark matter}\label{app::sidm_cross-section}
Since DM self-interaction is realized due to the presence of the term like $g_{d} X_\mu \overline{\chi}\gamma^\mu \chi$, the non-relativistic DM self-scattering can be well understood in terms of the attractive Yukawa potential
 \begin{equation}
     V(r)=\pm \frac{g_{d}^2}{4\pi r}e^{-M_{X}r}
 \end{equation}
where the $+(-)$ sign denotes repulsive (attractive) potential.\\
\\
To capture the relevant physics of forward scattering, the transfer cross-section is defined as
 \begin{equation*}
     \sigma_T=\int d\Omega (1-\cos\theta)\frac{d\sigma}{d\Omega}
 \end{equation*}
 In the Born limit, ${g_{d}}^2 m_{\chi}/(4\pi {M_{X}})\ll 1$,

\begin{equation*}
     \sigma^{\rm Born}_T=\frac{{g_{d}}^4}{2\pi {m_{\chi}^2} v_{\rm rel}^4} \left[ \log\left( 1+\frac{m^2_{\chi} v_{\rm rel}^2}{M_{X}^2} \right)-\frac{m_{\chi}^2 v_{\rm rel}^2}{M_{X}^2+m_{\chi}^2 v_{\rm rel}^2} \right]
\end{equation*}

Outside the Born limit, where   ${g_{d}}^2 m_{\chi}/(4\pi {M_{X}})\geq 1$, there can be two different regions: classical regime and resonance regime. In the classical regime (${m_{\chi}v_{\rm rel}}/{M_{X}}\geq 1$), solution for an attractive potential is given by \cite{Tulin:2012wi, Tulin:2013teo, PhysRevLett.90.225002}

\begin{equation*}
   \sigma^{\rm class.}_T=\begin{cases}
     \frac{4\pi}{M_{X}^2}\beta^2 \ln(1+\beta^{-1}) & \textbf{$\beta \leq 10^{-1}$}\\
     \frac{8\pi}{M_{X}^2}\left[ {\beta^2/(1+1.5\beta^{1.65})} \right] & \textbf{$10^{-1} < \beta \leq 10^3$} \\
     \frac{\pi}{M_{X}^2}\left[ \ln\beta +1-1/2 \ln^{-1}{\beta}\right]^2 & \textbf{$\beta \geq 10^3$}
   \end{cases}
\end{equation*}
and for the repulsive potential
\begin{equation*}
 	\sigma^{\rm class.}_T=\begin{cases}
 		\frac{2\pi}{M_{X}^2}\beta^2 \ln(1+\beta^{-2}) & \textbf{$\beta \leq 1$}\\
 		\frac{\pi}{M_{X}^2}\left[ {{\rm ln}(2 \beta) -{\rm ln}({\rm ln}2 \beta)} \right]^2 & \textbf{$ \beta \geq 1$} 
 	\end{cases}
\end{equation*}
where $\beta=\frac{2g_{d}^2{M_{X}}}{4\pi {m_{\chi}}v_{\rm rel}^2}$. 

Finally in the resonance region (${m_{\chi}v_{\rm rel}}/{M_{X}}\leq 1$), no analytical formula for $\sigma_T$ is available. So approximating the Yukawa potential by Hulthen potential $\left(V(r)=\pm \frac{{g_{d}}^2}{4\pi}\frac{\delta e^{-\delta r}}{1-e^{-\delta r}}\right)$, the transfer cross-section is obtained to be: 

\begin{equation*}
         \sigma_T^{\rm Hulthen}=\frac{16\pi \sin^2\delta_0}{m^2_\chi v_{\rm rel}^2}
\end{equation*}
 where $l=0$ phase shift $\delta_0$ is given by:
 $$ \delta_0=Arg \left[ \frac{i\Gamma(im_{\chi}v_{\rm rel}/\kappa M_{X})}{\Gamma(\lambda_+)\Gamma(\lambda_-)} \right]$$
 {\rm with}
 $$\lambda_{\pm}=1+\frac{im_{\chi}v_{\rm rel}}{2 \kappa M_{X}}\pm i^r\sqrt{\frac{{g_{d}}^2 m_{\chi}}{4\pi \kappa M_{X}}+(-1)^{r+1}\frac{m_{\chi}^2{v_{\rm rel}^2}}{4 \kappa^2 {M^2_{X}}}}$$
 
 and $\kappa \approx 1.6$ is a dimensionless number and $r=0(1)$ for attractive(repulsive) potential.

\section{Direct-detection constraints in the $U(1)_X$ model}\label{ddu1x}
Fig.~\ref{fig:S1} shows direct detection constraints from the DAMIC-M and DarkSide-50 experiments, for DM-electron scattering, indicated by the blue and green contours, respectively. We also explore the parameter space ${m_\chi, g_d}$ consistent with the observed relic density, using $M_X = 1$ MeV and $\epsilon = 10^{-9}$. The resulting cross-sections are shown by the magenta line, with the overlaid purple segment denoting regions that where it is possible to achieve sufficient DM self-interactions to address the small-scale structure problems of $\Lambda$CDM. This illustrates that the relic density–favored parameter space lies well beyond the sensitivity of current direct detection experiments.

%\newpage
%	\twocolumngrid
%	\bibliographystyle{JHEP}
\bibliographystyle{apsrev}
\bibstyle{apsrev}
%	\bibliography{ref,refn,ref1,ref12,ref3,ref4,ref5}	

\providecommand{\href}[2]{#2}\begingroup\raggedright\endgroup

\end{document}